\documentclass[preprintnumbers,nofootinbib,superscriptaddress,10pt]{revtex4}
\usepackage[dvipdfmx]{graphicx} 
\usepackage{amsmath,amssymb,amsfonts,bm} 

\def\a{\alpha}  \def\g{\gamma}  \def\d{\delta}          \def\m{\mu} \def\n{\nu}     \def\r{\rho}   \def\t{\tau}       

\def\dg{\dagger}  \def\nn{\nonumber}

    \newcommand{\To}{\Rightarrow}

\def\abs#1{\left| #1\right|}

\newcommand{\Diag}[3]{ \begin{pmatrix} #1 & 0 & 0 \\ 0 & #2 & 0 \\ 0 & 0 & #3 \\\end{pmatrix}}

\usepackage{color}

\begin{document}

\title{\large Explicit rephasing  to Kobayashi--Maskawa representation and \\
fundamental phase structure of CP violation}

\preprint{STUPP-25-291}

\author{Masaki J. S. Yang}
\email{mjsyang@mail.saitama-u.ac.jp}
\affiliation{Department of Physics, Saitama University, 
Shimo-okubo, Sakura-ku, Saitama, 338-8570, Japan}
\affiliation{Department of Physics, Graduate School of Engineering Science,
Yokohama National University, Yokohama, 240-8501, Japan}



\begin{abstract} 

In this letter, we construct an explicit rephasing transformation that converts an arbitrary unitary matrix into the Kobayashi--Maskawa (KM) parameterization and identify all independent CP phases in the mixing matrix as the arguments of its matrix elements.
Furthermore, by applying this rephasing transformation to the fermion diagonalization matrices $U^{\n, e}$, we show that the Majorana phases are represented by  fermion-specific phases $\delta^{\nu, e}_{\rm KM}$ and their relative phases.
In particular, by neglecting the 3-1 elements $U_{31}^{\nu ,e}$,  the KM phase $\delta_{\rm KM}$ is concisely expressed as $\delta_{\rm KM}  = \arg \left [1 + ({U^{e * }_{21} U^{\nu}_{21}  / U^{e * }_{11}  U^{\nu}_{11} }) \right ] 
+ \arg \left [ - { U_{32}^{e *}  U_{32}^{\nu} / U^{e * }_{22} U^{\nu}_{22}  } \right] $ by fermion-specific rephasing invariants representing two relative phases. 

\end{abstract} 

\maketitle

\section{Introduction}

The CP phases appearing in mixing matrices provide the most direct representation of flavored CP violation in the Standard Model. 
Beginning with the Jarlskog invariant~\cite{Jarlskog:1985ht}, 
a wide variety of rephasing invariants have been proposed and extensively studied in both quark and lepton mixing  
\cite{Jarlskog:1985ht, Wu:1985ea, Bernabeu:1986fc, Gronau:1986xb, Branco:1987mj, Bjorken:1987tr, Nieves:1987pp, Botella:1994cs, Kuo:2005pf, Jenkins:2007ip, Chiu:2015ega,  Denton:2020igp}. 
Many of these quantities are designed to capture only the magnitude of CP violation, typically through imaginary parts or $\sin\delta$-type expressions. However, they are not sufficient to identify the phase structure of the mixing matrix itself.
In particular, although many papers have declared the ``rephasing'' of an arbitrary unitary matrix into the standard parametrization, the specific procedure to realize this transformation has largely been treated implicitly. 

In recent years, a new class of rephasing invariants involving the determinant of the mixing matrices has been systematically studied \cite{Yang:2025hex,Yang:2025cya,Yang:2025law,Yang:2025ftl,Yang:2025dhm,Yang:2025vrs},  leading to compact formulae of CP phases. 
By extending these results to Majorana phases and other unphysical phases, an explicit rephasing transformation into the PDG parametrization has been obtained \cite{Yang:2025dkm, Yang:2025qlg}. 

In this work, we further advance this line of research by constructing an explicit rephasing transformation in the original Kobayashi--Maskawa (KM) parametrization~\cite{Kobayashi:1973fv}.
It provides complete rephasing matrices that transform an arbitrary unitary matrix into the KM form, expressed solely in terms of the arguments of the mixing-matrix elements. 

Moreover, the rephasing transformation is also applied to the diagonalization matrices $U^{\n, e}$ of individual fermions.
This allows the observed CP phases to be understood as combinations of more fundamental fermion-specific phases and their relative phases, providing a systematic framework for analyzing the origin of CP violation.
While the CP phase in the KM parametrization has been discussed extensively \cite{Koide:2004gj, Koide:2008yu, Frampton:2010ii,Dueck:2010fa,Frampton:2010uq,Li:2010ae,Qin:2011bq,Zhou:2011xm,Qin:2011ub,Qin:2010hn,Zhang:2012ys,Li:2012zxa,Zhang:2012bk}, such a direct representation of the phases will present various improvements for both future experimental and theoretical studies. 

\section{Explicit rephasing transformation to KM parametrization}

Here, we present an explicit rephasing transformation to the Kobayashi--Maskawa parametrization $U^{1}$ for an arbitrary phase convention, by deriving the KM phase $\delta_{\rm KM}$ and the Majorana phases $\alpha_{2,3}$ in this parameterization.  
Let us define the KM form $U^{1}$ and the phase matrix $P$ with the Majorana phases $\alpha_i$, 
\begin{align}
U^{1} P
& \equiv 
\begin{pmatrix}
1 & 0 & 0 \\
0 & c_{2} & -s_{2} \\
0 & s_{2} & c_{2}
\end{pmatrix}
\begin{pmatrix}
c_1 & -s_1 & 0 \\
s_1 & c_1 & 0 \\
0 & 0 & e^{i\delta_{\rm KM}}
\end{pmatrix} 
\begin{pmatrix}
1 & 0 & 0 \\
0 & c_{3} & s_{3} \\
0 & s_{3} & - c_{3}
\end{pmatrix} P  \label{KMrep} \nn \\
& \equiv 
\begin{pmatrix}
c_1 & -s_1 c_3 & -s_1 s_3 \\
s_1 c_2 & c_1 c_2 c_3 - s_2 s_3 e^{i\delta_{\rm KM}} & c_1 c_2 s_3 + s_2 c_3 e^{i\delta_{\rm KM}} \\
s_1 s_2 & c_1 s_2 c_3 + c_2 s_3 e^{i\delta_{\rm KM}} & c_1 s_2 s_3 - c_2 c_3 e^{i\delta_{\rm KM}}
\end{pmatrix} 
\Diag{1}{e^{ i \a_{2} / 2}}{e^{ i \a_{3} / 2}}  . 
\end{align}
As in the PDG parametrization, one can choose all mixing angles to be positive $c_i, s_i > 0$ by phase redefinitions of five fermion fields and the sign freedom of $\delta_{\rm KM}$.
The following six conditions specify the phase structure of this parametrization
\begin{align}
\arg U_{e1}^{1} = \arg [- U_{e 2}^{1}] = \arg [- U_{e3}^{1}] = \arg U_{\m 1}^{1} = \arg U_{\t 1}^{1} = 0 \, , ~~ 
\arg  \det U^{1} = \delta_{\rm KM} + \pi \, .
\end{align}

The rephasing transformation from $U^{1}$ to a mixing matrix defined in an arbitrary basis $U$ is written by 
\begin{align}
\begin{pmatrix}
U_{e1} & U_{e2} & U_{e3}  \\
U_{\m1} & U_{\m 2} & U_{\m 3}  \\
U_{\t 1} & U_{\t 2} & U_{\t 3}  \\
\end{pmatrix}
= 
\Diag{e^{i \g_{L1}}}{e^{i \g_{L2}}}{e^{i \g_{L3}}}
\begin{pmatrix}
|U_{e1}| & - |U_{e2}| &- |U_{e3}| \\
|U_{\m1}| & U_{\m 2}^{1} & U_{\m 3}^{1}  \\
|U_{\t 1}| & U_{\t 2}^{1} & U_{\t 3}^{1} \\
\end{pmatrix}
\Diag{e^{ - i \g_{R1}}}{e^{ - i \g_{R2}}}{e^{ - i \g_{R3}}} . 
\end{align}
Removing six phases $\g_{Li} , \g_{Ri}$ by the determinant and matrix elements carrying trivial arguments, 
the KM phase $\delta_{\rm KM}$ is obtained as
\begin{align}
\arg \left[ {U_{e2} U_{e3} U_{\m1} U_{\t 1} } \right]
 & = 2 \g_{L1}  + \g_{L2} + \g_{L3} - 2 \g_{R1}  - \g_{R2} - \g_{R3}  \, , \nn \\
 \arg \left[- {U_{e2} U_{e3} U_{\m1} U_{\t 1} \over  \det U } \right] & = 
 \g_{L1}  -  \g_{R1} - \d_{\rm KM}
 ~~ \To ~~ 
\d_{\rm KM} = \arg \left[ - {  U_{e1} \det U \over U_{e2} U_{e3} U_{\m1} U_{\t 1}  } \right]  . 
\end{align}
Such a simple and important calculation had not been demonstrated in more than half a century since the original work. 

This KM phase is observable because 
it is a sum of the PDG phase and the angles of the unitarity triangle $\a, \g$ by the sum rule~\cite{Yang:2025ftl} 
\begin{align}
  \d_{\rm KM}  &= \pi - \d_{\rm PDG} - \a + \g    \, , \nn \\
 \arg \left[ - {  U_{e 1} \det U \over U_{e 2} U_{e 3} U_{\m 1} U_{\t 1}  } \right]
&=  \arg \left[ - \frac{ U_{e 3} \det U }{ U_{e1} U_{e2} U_{\m 3} U_{\t 3} } \right]
 - \arg \left [ - { U_{\t 1}^{} U_{\t 3}^{*} \over U_{e 1}^{} U_{e 3}^{*} } \right ]
+ \arg \left [ - { U_{e 1}^{} U_{e 3}^{*} \over U_{\m 1}^{} U_{\m 3}^{*} }  \right ]  \, .   
\end{align}
Without resorting to the unitarity triangle, 
one can directly evaluate one phase from the other parametrization.
\begin{align}
\d_{\rm KM}  = \d_{\rm PDG} + \arg
\left[ { s_{12} c_{23} + c_{12} s_{23} s_{13}  e^{- i \d_{\rm PDG}} \over s_{12} s_{23} - c_{12} c_{23} s_{13} e^{i \d_{\rm PDG}} } \right] , ~~~ 
\d_{\rm PDG}  =   \arg \left[ - {c_1 c_2 s_3 + s_2 c_3 e^{ i\delta_{\rm KM}} \over c_1 s_2 s_3 - c_2 c_3 e^{- i\delta_{\rm KM}}} \right] - \d_{\rm KM} , 
\end{align}
where $s_{ij}$ and $c_{ij}$ denote the mixing angles in the PDG parametrization. 

As in the previous letter~\cite{Yang:2025dkm}, we perform an explicit rephasing transformation by solving for the phases $\gamma_{L i}$ and $\gamma_{R i}$.
The five conditions to be solved are 
\begin{align}
\g_{L1} - \g_{R1} & = \arg U_{e1} \, , ~ 
\g_{L1} - \g_{R2} = \arg[ - U_{e2}] \, , ~ 
\g_{L1} - \g_{R3} = \arg [- U_{e3}] \, , \nn \\
\g_{L2} - \g_{R1} & = \arg U_{\m 1} \, , ~ 
\g_{L3} - \g_{R1} = \arg U_{\t 1} \, . 
\end{align}
Because of the freedom associated with an overall phase, one phase necessarily remains undetermined.
By choosing $\gamma_{R1}$ as the unsolved variable to preserve the structure of Majorana phases, 
$\gamma_{R1}$ cancels between the left and right phase matrices.
As a result, the explicit rephasing transformation is found to be
\begin{align}
U =  
\Diag{e^{i \arg U_{e1} }}{e^{i  \arg U_{\m 1} }}{e^{i \arg U_{\t 1} }}
U^{1}
\Diag{1 }{e^{ i \arg [- U_{e2}/  U_{e1}] }}{e^{  i \arg [- U_{e3}/ U_{e1}] }} .
\end{align}
We find that this rephasing transformation ultimately amounts to trivializing five phases. 
In this parametrization, $\gamma_{R2}$ and $\gamma_{R3}$ directly yield the Majorana phases 
\begin{align}
{\a_{2} \over 2} =  \arg \left [- {U_{e2} \over  U_{e1}} \right] \, , ~~~ 
{\a_{3} \over 2} =  \arg \left[- {U_{e3} \over  U_{e1}} \right] \, .
\end{align}
Unlike in the PDG parametrization, the phase $\alpha_{3}/2$ is not accompanied by $\delta_{\rm KM}$.

With this explicit rephasing, the independent six phases of a unitary matrix
 are expressed in terms of arguments of the matrix elements and the determinant.
Consequently, the remaining nontrivial arguments are written in terms of these six phases.
Performing the rephasing transformation, we obtain
\begin{align}
U & =  
\Diag{e^{i \arg U_{e1} }}{e^{i  \arg U_{\m 1} }}{e^{i \arg U_{\t 1} }}
\begin{pmatrix}
|U_{e1}| & - |U_{e2}| &- |U_{e3}| \\
|U_{\m1}| & U_{\m 2}^{1} & U_{\m 3}^{1}  \\
|U_{\t 1}| & U_{\t 2}^{1} & U_{\t 3}^{1} \\
\end{pmatrix}
\Diag{1 }{e^{ i \arg [- U_{e2}/  U_{e1}] }}{e^{  i \arg [- U_{e3}/ U_{e1}] }} \nn \\
& = 
\begin{pmatrix}
U_{e1} & U_{e2} & U_{e3} \\
U_{\m1} & e^{ i \arg [- U_{e2}U_{\m 1} / U_{e1}] } U_{\m 2}^{1} &  e^{  i \arg [- U_{e3} U_{\m 1}/ U_{e1}] }U_{\m 3}^{1}  \\
U_{\t 1} & e^{ i \arg [- U_{e2} U_{\t 1} / U_{e1}] } U_{\t 2}^{1} &  e^{  i \arg [- U_{e3}U_{\t 1} / U_{e1}] } U_{\t 3}^{1} \\
\end{pmatrix} . 
\end{align}
From this, the remaining four nontrivial arguments are written precisely by the angles of unitarity triangles 
\begin{align}
\arg U_{\m 2}^{1} & = \arg \left[ - {  U_{e 1}U_{\m2}  \over U_{e2} U_{\m1}  } \right] \, , ~~ 
\arg U_{\m 3}^{1} = \arg \left[ - { U_{e 1} U_{\m3}  \over U_{e3} U_{\m1}  } \right] \, , \nn \\
\arg U_{\t 2}^{1} & = \arg \left[ - { U_{e 1} U_{\t 2} \over U_{e2} U_{\t 1} } \right] \, , ~~ 
\arg U_{\t 3}^{1}  = \arg \left[ - { U_{e 1} U_{\t 3}  \over U_{e3} U_{\t1}  } \right] \, . 
\end{align}
Since $U_{e1}, U_{e2}, U_{e3}, U_{\m 1}$ and $U_{\t 1}$  carry no arguments  in the KM parametrization, 
the validity of relations is evident from the rephasing invariance of the angles.

The rephasing transformation between the PDG and KM parametrizations identifies the relation between the Majorana phases and organizes the correspondence between the two representations.
The explicit rephasing to the PDG parameterization $U^{0}$ from an arbitrary  mixing matrix $U$ is given by
\begin{align}
U^{0} = 
\Diag{e^{ - i \arg U_{e1} } }{e^{ i  \arg \left[ {U_{e2}  U_{\t 3} \over \det U  } \right]}}{e^{ i \arg \left[ {  U_{e2} U_{\m3} \over \det U } \right] }}
U 
\Diag{1}{e^{ - i \arg [{U_{e2}\over U_{e1} }] } }{e^{ - i \arg \left[ { U_{e2} U_{\m3} U_{\t 3} \over \det U } \right] }} . 
\end{align}
From this, the Majorana and the Dirac phases in the PDG parametrization are 
\begin{align}
{\a_{2}^{\rm PDG} \over 2} =  \arg \left [ {U_{e2} \over  U_{e1}} \right] \, , ~~
{\a_{3}^{\rm PDG} \over 2} =  \arg \left[ { U_{e2} U_{\m3} U_{\t 3} \over \det U } \right] \, , ~~ 
\d_{\rm PDG} = \arg \left[ { U_{e1} U_{e2} U_{\m3} U_{\t 3} \over U_{e3} \det U } \right] \, .
\end{align}
Combining these two explicit transformations, we obtain
\begin{align}
U^{0} = \Diag{1}{e^{ i  \arg \left[ {U_{e2} U_{\m 1}  U_{\t 3} \over \det U  } \right]}}{e^{ i \arg \left[ {  U_{e2} U_{\m3} U_{\t 1} \over \det U } \right] }} 
U^{1} 
\Diag{1 }{-1}{ - e^{ - i  \d _{\rm PDG} } }  \, . 
\end{align}
The left-handed rephasing transformation is expressed in terms of third-order invariants because  they correspond to the arguments of $U_{\m1}^{0}$ and $U_{\t 1}^{0}$ in the first column of $U^{0}$. 
As a result, the relations of the Majorana phases in the two parametrizations are found to be
\begin{align}
{\a_{2} \over 2} = {\a_{2}^{\rm PDG} \over 2}  + \pi \, , ~~~ 
{\a_{3} \over 2} = {\a_{3}^{\rm PDG} \over 2} - \d_{\rm PDG} + \pi \, . 
\end{align}
%
Such simple relations will hold between other parametrizations.

\section{Expressions of CP phases by fermion-specific rephasing invariants for simplified mixing matrix }

The explicit rephasing transformation can be applied not only to the mixing matrix itself but also to 
the diagonalization matrices of neutrinos $U^{\nu}$ and charged leptons $U^{e}$. 
%
We define explicit rephasing transformations 
$U^{\nu,e} = \Phi^{\nu,e}_{L} \, U^{\nu,e 1} \, \Phi^{\nu,e}_{R},$
where $U^{\nu,e 1}$ are their KM parametrization, and 
$\Phi^{\nu,e}_{L,R}$ are phase matrices expressed in terms of the corresponding matrix elements.
Then the lepton mixing matrix $U_{l} \equiv U^{e \dagger} U^{\nu}$ is rewritten as 
\begin{align}
U_{l} = \Phi^{e \dagger}_{R} \, U^{e 1 \dagger} \, \Phi_L^{e\dagger} \Phi_L^\nu \, U^{\nu 1} \, \Phi_R^{\n} \, .
\end{align}

The first candidates of CP-violating sources are fermion-specific 
 KM phases $\d^{\nu,e}_{\rm KM}$ written in rephasing invariants; 
\begin{align}
\d^{\n,e}_{\rm KM} \equiv  \arg \left[ -  {U_{11}^{\n , e} \det U^{\n , e} \over U_{12}^{\n , e} U_{13}^{\n , e} U_{21}^{\n , e} U_{31}^{\n , e}  } \right] \, . 
\end{align}
Among the remaining CP phases, the left-handed rephasing also affects physical CP violation.
By combining the left-handed phases as $\Phi^L \equiv \Phi_e^{L\dagger} \Phi_\nu^L \equiv {\rm diag} \, ( e^{i\r_{1}} \, , \, e^{i\r_{2}} \, , \, e^{i \r_{3}})$, 
the relative phases between $U^{\nu 1}$ and $U^{e 1}$ are given by 
\begin{align}
\r_{1} = \arg [U^{e * }_{11} U^{\n}_{11} ] \, , ~
\r_{2} = \arg [U^{e *}_{21} U^{\n}_{21} ] \, , ~ 
\r_{3} = \arg [U^{e *}_{31} U^{\n}_{31} ] \, .
\end{align}
Only the relative phases $\rho_i-\rho_j$ are physically meaningful  
\begin{align}
\r_{1} - \r_{2} = \arg \left[ { U^{e * }_{11} U^{\n}_{11} \over U^{e *}_{21} U^{\n}_{21} } \right] \, , ~ 
\r_{2} - \r_{3} = \arg \left[ { U^{e *}_{21} U^{\n}_{21} \over U^{e *}_{31} U^{\n}_{31} } \right] \, , ~ 
\r_{3} - \r_{1} = \arg \left[ { U^{e *}_{31} U^{\n}_{31} \over U^{e * }_{11} U^{\n}_{11} } \right] \, . ~ 
\label{relatives}
\end{align}
These relative phases are manifestly invariant under rephasing of $U_{l}$, 
and two of $\rho_i-\rho_j$ are independent.

\subsection{Majorana phases}

The Majorana phases $\a_{2,3}$ are functions of $\Phi^{\nu}_{R}$, $\delta_{\rm KM}^{\nu}$, 
and the relative CP phases $\r_{i} - \r_{j}$, because the matrix $\Phi^{e}_{R}$ and the phase $\delta_{\rm KM}^{e}$ are irrelevant. 
To express these phases, we expand the matrix element $U_{e1}$ as
\begin{align}
U_{e1} &= U^{e *}_{11} U^{\n}_{11} + 
U^{e *}_{21} U^{\n}_{21} + U^{e *}_{31} U^{\n}_{31}  \nn \\
&= e^{ i \arg [U^{e *}_{11} U^{\n}_{11}]} 
( \abs {U^{e}_{11} U^{\n}_{11}} + e^{ i \arg [{  U^{e *}_{21} U^{\n}_{21} \over U^{e *}_{11}  U^{\n}_{11}}] } \abs {U^{e}_{21} U^{\n}_{21}} + 
e^{ i \arg [{ U^{e *}_{31} U^{\n}_{31} \over U^{e* }_{11}  U^{\n}_{11}}] } \abs {U^{e}_{31} U^{\n}_{31}} ) \, . 
\end{align}
Similarly, the matrix element $U_{e2}$ is 
\begin{align}
U_{e2} 
&= e^{ i \arg [U^{e *}_{11} U^{\n}_{12}]} ( \abs {U^{e}_{11} U^{\n}_{12}} + 
e^{ i \arg [  { U^{e *}_{21} \over U^{e *}_{11} U^{\n}_{12} } \cdot  - { U^{\n}_{12} U^{\n}_{21} \over U^{\n}_{11} }]}  |U^{e}_{21}| U^{\n 1}_{22} + 
e^{ i \arg [{ U^{e *}_{31} \over U^{e *}_{11} U^{\n}_{12} } \cdot - { U^{\n}_{12} U^{\n}_{31} \over U^{\n}_{11} }]} 
|U^{e}_{31}| U^{\n 1}_{32} ) \, . 
\end{align}
The nontrivial arguments $\arg U^{\nu 1}_{22}$ and $\arg U^{\nu 1}_{32}$ of the KM parametrization $U^{\nu 1}$ are  functions of $\delta_{\rm KM}^{\nu}$ and the mixing angles by the unitarity. 
The Majorana phases are expressed as
\begin{align}
{\a_{2} \over 2} = \arg \left[ - {U_{e2} \over U_{e1}} \right]
& = \arg \left [- { U^{\n}_{12} \over U^{\n}_{11}} \right ]
 + \arg \left[  
 \frac{ \abs {U^{e}_{11} U^{\n}_{12}} - e^{ i (\r_{2} - \r_{1})} |U^{e}_{21}| U^{\n 1}_{22} - e^{ i  (\r_{3}-\r_{1} )} |U^{e}_{31}| U^{\n 1}_{32}}
 {\abs {U^{e}_{11} U^{\n}_{11}} + e^{ i (\r_{2} - \r_{1}) } |U^{e}_{21} U^{\n}_{21}| + e^{ i  (\r_{3} - \r_{1} )} |U^{e}_{31} U^{\n}_{31}| } \right] \, . 
\end{align}
To illustrate the advantage of this representation, we also consider the Majorana phase 
represented by PDG parameterizations $U^{\n 0}$ and $U^{e 0}$ of diagonalization matrices $U^{\n,e}$, 
\begin{align}
{\a_{2}^{\rm PDG} \over 2} & = \arg \left [{ U^{\n}_{12} \over U^{\n}_{11}} \right ]
 + \arg \left[ \frac{ |U^{e}_{11} U^{\n}_{12}| + e^{ i (\r'_{2} - \r'_{1})} U^{e 0 *}_{21} U^{\n 0}_{22} 
 + e^{ i (\r'_{3} - \r'_{1})} U^{e 0 *}_{31} U^{\n 0}_{32} }
 {|U^{e}_{11} U^{\n}_{11}| + e^{ i (\r'_{2} - \r'_{1}) } U^{e 0 *}_{21} U^{\n 0}_{21} + e^{ i (\r'_{3} - \r'_{1}) } U^{e 0 *}_{31} U^{\n 0}_{31}  } \right] . 
\end{align}
Here, $\r'_{1,2,3}$ denote the relative phases between the PDG forms, and $U^{e0}_{21}$ and $U^{e0}_{31}$ contain the nontrivial phase $\d^{e}_{\rm PDG}$. 
Compared to the PDG forms, the elements $U^{e}_{21}$ and $U^{e}_{31}$ in the KM representation have trivial phases. 
Since one of the CP phases of the charged leptons becomes irrelevant, the KM form is one of the most suitable parameterizations 
for describing Majorana phases. 

Finally, the matrix element $U_{e3}$ is 
\begin{align}
U_{e3} 
&= e^{ i \arg [U^{e *}_{11} U^{\n}_{13}]} (\abs {U^{e}_{11} U^{\n}_{13}} + 
e^{i \arg [{U^{e *}_{21} \over U^{e *}_{11} U^{\n}_{13}} \cdot  - { U^{\n}_{13} U^{\n}_{21} \over U^{\n}_{11} }]}   | U^{e}_{21}| U^{\n 1}_{23} + 
e^{i \arg [{U^{e *}_{31} \over U^{e *}_{11} U^{\n}_{13}} \cdot - { U^{\n}_{13} U^{\n}_{31} \over U^{\n}_{11} }]}  |U^{e}_{31}|  U^{\n 1}_{33}  ) \, , 
\end{align}
and the other Majorana phase is obtained as
\begin{align}
{\a_{3} \over 2}  &= \arg \left[- { U_{e3} \over U_{e1} } \right] 
 = \arg \left[- {U_{13}^{\n} \over U_{11}^{\n } } \right ] + 
\arg  \left[ {  \abs {U^{e}_{11} U^{\n}_{13}} - e^{ i (\r_{2} - \r_{1} )} |U^{e}_{21}| U^{\n 1}_{23} - e^{ i  (\r_{3}-\r_{1} )} |U^{e}_{31}| U^{\n 1}_{33} 
  \over 
|U^{e}_{11} U^{\n}_{11}| + e^{ i (\r_{2} - \r_{1}) } |U^{e}_{21} U^{\n}_{21}| + e^{ i  (\r_{3} - \r_{1} )} |U^{e}_{31} U^{\n}_{31}| } 
 \right]  . 
\end{align}

If the mixings of $U^{e}$ are as small as the CKM matrix due to some grand unified relation,
the third term involving $U^{e}_{31}$ is negligible for $O(1)$ values of $U^{\nu}_{ij}$. 
As a result, the Majorana phases are approximated as  
\begin{align}
{\a_{2} \over 2} & \simeq \arg \left [- { U^{\n}_{12} \over U^{\n}_{11}} \right ]
 + \arg \left[  
 \frac{  \abs {U^{e}_{11} U^{\n}_{12}} - e^{ i (\r_{2} - \r_{1})} |U^{e}_{21}| U^{\n 1}_{22} }
 {\abs {U^{e}_{11} U^{\n}_{11}} + e^{ i (\r_{2} - \r_{1}) } |U^{e}_{21} U^{\n}_{21}| } \right] \, , \\
 {\a_{3} \over 2}  & \simeq \arg \left[ - {U_{13}^{\n} \over U_{11}^{\n } } \right ] + 
\arg  \left[ {  \abs {U^{e}_{11} U^{\n}_{13}} - e^{ i (\r_{2} - \r_{1})} |U^{e}_{21}| U^{\n 1}_{23}  
  \over 
|U^{e}_{11} U^{\n}_{11}| + e^{ i (\r_{2} - \r_{1}) } |U^{e}_{21} U^{\n}_{21}| } 
 \right] \, .
\end{align}
In this limit, the Majorana phases $\alpha_{2,3}$ reduce to functions of three phases:
the KM phase of neutrinos $\delta^{\nu}_{\rm KM}$, one relative phase, 
and one corresponding Majorana-like phase. 
However, because its original definition contains
 superficial sign factors, it is more practical to remove them by appropriate redefinitions. 

\subsection{Simplified expression for KM phase}

Since the approximation $U^{\nu,e}_{31}=0$ removes the fermion-specific 
phases $\delta^{\nu , e}_{\rm KM}$, the KM phase $\d_{\rm KM}$ depends solely on two relative phases.
To demonstrate this behavior, we impose $U^{\nu,e}_{31}=0$ and use the inversion formula 
partially for matrix elements with nontrivial arguments,
\begin{align}
U_{l} = U^{e \dg} U^{\n} & =
\begin{pmatrix}
 U^{e * }_{11}  &  U^{e * }_{21}  & 0 \\[2pt]
- {U_{21}^{e } U_{33}^{e} \over \det U^{e}} & {U_{11}^{e} U_{33}^{e} \over \det U^{e}} & U^{e *}_{32} \\[2pt]
{U_{21}^{e} U_{32}^{e} \over \det U^{e} } & - {U_{11}^{e} U_{32}^{e} \over \det U^{e} } & U^{e * }_{33} \\
\end{pmatrix}
\begin{pmatrix}
 U^{\n}_{11}  & - {U_{21}^{\n *} U_{33}^{\n *} \over \det U^{\n *}} & {U_{21}^{\n *} U_{32}^{\n *} \over \det U^{\n *} } \\[2pt]
 U^{\n}_{21}  & {U_{11}^{\n *} U_{33}^{\n *} \over \det U^{\n *}} & - {U_{11}^{\n *} U_{32}^{\n *} \over \det U^{\n *} } \\[2pt]
0 & U^{\n}_{32}  & U^{\n}_{33} \\
\end{pmatrix} \nn 
\\ & = 
\begin{pmatrix}
 U^{e * }_{11} U^{\n}_{11} + U^{e * }_{21} U^{\n}_{21} & {U_{33}^{\n *} \over \det U^{\n *}} ( U^{e * }_{21} U_{11}^{\n *} - U^{e * }_{11} U_{21}^{\n *}  ) & {U_{32}^{\n *} \over \det U^{\n *} } (U^{e * }_{11} U_{21}^{\n *}  -  U^{e * }_{21} U_{11}^{\n *} ) \\[2pt]
 {U_{33}^{e} \over \det U^{e}} (U_{11}^{e}  U^{\n}_{21}  - {U_{21}^{e } } U^{\n}_{11} )& * & * \\[2pt]
{U_{32}^{e} \over \det U^{e} } (U_{21}^{e} U^{\n}_{11} - U_{11}^{e} U^{\n}_{21} )& * & * \\
\end{pmatrix} . 
\end{align}
Here, the matrix elements denoted by $*$ are not important in the phase calculation.

Since the four matrix elements carry mutually complex-conjugate factors, 
the KM phase in this approximation is concisely expressed as
\begin{align}
\d_{\rm KM} & =  \arg \left [ - { (U^{e * }_{11} U^{\n}_{11} + U^{e * }_{21} U^{\n}_{21} ) \det U^{e} \det U^{\n *} \over  U_{32}^{e} U_{33}^{e} U_{32}^{\n *}  U_{33}^{\n *} } \right] \nn  \\
 & = \arg \left  [1 + {U^{e * }_{21} U^{\n}_{21}  \over U^{e * }_{11}  U^{\n}_{11} }  \right ] 
 + \arg \left [ - {\det U^{e} \over U^{e }_{11} U_{32}^{e} U_{33}^{e} }  { \det U^{\n *} \over U^{\n *}_{11} U_{32}^{\n *}  U_{33}^{\n *}  }  \right]\, . 
\end{align}

To see how the two relative phases are separated, we further consider the approximation in which $U^e_{21}$ is neglected.
Rewriting the second term without invoking the inversion formula, 
\begin{align}
\d_{\rm KM} \simeq  \arg \left [ - {\det U^{e} \over U^{e }_{11} U_{32}^{e} U_{33}^{e} }  { \det U^{\n *} \over U^{\n *}_{11} U_{32}^{\n *}  U_{33}^{\n *}  }  \right]
= \arg \left [ - { U_{32}^{e *}  U_{32}^{\n} \over U^{e * }_{22} U^{\n}_{22}  } \right]
= \arg \left [- { U_{33}^{e*} U_{33}^{\n}  \over U^{e * }_{23} U^{\n}_{23}  } \right] 
=  \r_{3} - \r_{2} + \pi  \, .  
\label{simpleKM}
\end{align}
This relation holds in place of Eq.~(\ref{relatives}) in the situation where $U^{\nu,e}_{31} = 0$ and 
all CP phases can be removed by a rephasing. 
This result (\ref{simpleKM})  has a simple interpretation. 
If $U^{e}_{21}$, $U^{e}_{31}$, and $U^{\nu}_{31}$ are set to zero,
the mixing matrix is cast into the following form after suitable phase redefinitions,
\begin{align}
\begin{pmatrix}
1 & 0 & 0 \\
0 & c_{23}^{e} & -s_{23}^{e} \\
0 & s_{23}^{e} & c_{23}^{e}
\end{pmatrix}
\Diag{e^{ i \r_{1} }}{e^{ i \r_{2} }}{e^{ i \r_{3} }} 
\begin{pmatrix}
c_{12}^{\n} & - s_{12}^{\n} & 0 \\
s_{12}^{\n} & c_{12}^{\n} & 0 \\
0 & 0 & 1
\end{pmatrix} 
\begin{pmatrix}
1 & 0 & 0 \\
0 & c_{23}^{\n} & s_{23}^{\n} \\
0 & s_{23}^{\n} & - c_{23}^{\n}
\end{pmatrix} .
\end{align}
The result $\delta_{\rm KM} = \rho_{3} - \rho_{2} + \pi$ corresponds to rephasing the matrix into the KM form~(\ref{KMrep}) by field redefinitions.

In the general case where $U^{e}_{21}$ is not neglected, the KM phase is rewritten without using the determinant
\begin{align}
\d_{\rm KM}  = \arg \left [1 + {U^{e * }_{21} U^{\n}_{21}  \over U^{e * }_{11}  U^{\n}_{11} } \right ] 
+ \arg \left [ - { U_{32}^{e *}  U_{32}^{\n} \over U^{e * }_{22} U^{\n}_{22}  } \right] \, . 
\end{align}
In a similar parameterization 
\begin{align}
\begin{pmatrix}
1 & 0 & 0 \\
0 & c_{23}^{e} & -s_{23}^{e} \\
0 & s_{23}^{e} & c_{23}^{e}
\end{pmatrix}
\begin{pmatrix}
c_{12}^{e} & - s_{12}^{e} & 0 \\
s_{12}^{e} & c_{12}^{e} & 0 \\
0 & 0 & 1
\end{pmatrix} 
\Diag{e^{ i \r_{1} }}{e^{ i \r_{2} }}{e^{ i \r_{3} }} 
\begin{pmatrix}
c_{12}^{\n} & - s_{12}^{\n} & 0 \\
s_{12}^{\n} & c_{12}^{\n} & 0 \\
0 & 0 & 1
\end{pmatrix} 
\begin{pmatrix}
1 & 0 & 0 \\
0 & c_{23}^{\n} & s_{23}^{\n} \\
0 & s_{23}^{\n} & - c_{23}^{\n}
\end{pmatrix} ,
\end{align}
the KM phase is obtained as 
\begin{align}
\d_{\rm KM} = \arg [ - e^{i (\rho _3 - \rho_{2})} ( c^e_{12} c^{\nu }_{12} - e^{i (\rho _2 - \rho _1)} s^e_{12} s^{\nu }_{12} )] \, . 
\end{align}
If the 2-1 element on the charged lepton mixing is small, 
the KM phase effectively reflects the relative phase of the heavier generations. 

Finally, let us  consider the limit in which the three off-diagonal elements of $U^{e}$
 (e.g., $U^{e}_{21}$, $U^{e}_{31}$, and $U^{e}_{32}$) are neglected. 
 In this limit, $U^{e}$ becomes a diagonal matrix by the unitarity, and it leads to $\arg [U^{e}_{11} U^{e}_{22} U^{e}_{33} / \det U^{e}] = 0$. 
 Since all the phase of $U^{e}$ are removed by a rephasing transformation, 
 the result of the formula coincides with the CP phase of $U_{l} = U^{\n}$,
\begin{align}
\d_{\rm KM} =  \arg \left[ -  {U_{11}^{\n} \det U^{\n} \over U_{12}^{\n} U_{13}^{\n} U_{21}^{\n} U_{31}^{\n}  } \right] =  \d_{\rm KM}^{\n} \, . 
\end{align}
In the above calculation, we set $U_{31}^{\n} = 0$, and thus $\d_{\rm KM} = 0$. 
Such calculations provide a benchmark for analyzing the rephasing-invariant behavior of CP phases in general mixing matrices with nonzero $U^{\nu, e}_{31}$ \cite{Yang:2024ulq, Yang:2025yst}.

\section{Summary}

In this letter, we construct an explicit rephasing transformation that maps an arbitrary unitary matrix to the Kobayashi--Maskawa parameterization, and identify all CP phases of the rephasing as arguments of matrix elements.
By applying the rephasing transformation to fermion diagonalization matrices $U^{\n, e}$, we further show that the observable CP phases are expressed in terms of fermion-specific KM phases $\d^{\n ,e}_{\rm KM}$ and relative phases between $U^{\n, e}$.
This analysis demonstrates that, in the description of Majorana phases, the KM parameterization involves fewer CP phases than the PDG parameterization and thus possesses practical advantages.

Moreover, under the approximation in which the 3-1 elements of $U^{\n, e}$ are neglected, 
the KM phase is expressed compactly in terms of only two relative phases. 
As a result, we obtain a remarkably simplified relation
$\d_{\rm KM}  = \arg \left [1 + ({U^{e * }_{21} U^{\n}_{21}  / U^{e * }_{11}  U^{\n}_{11} }) \right ] 
+ \arg \left [ - { U_{32}^{e *}  U_{32}^{\n} / U^{e * }_{22} U^{\n}_{22}  } \right]$.  
In this formulation, the CP violation is understood as a direct consequence of a more fundamental phase structure.

These representations of CP phases in terms of fermion-specific phases are 
applicable to a wide range of contexts,  such as 
$B$-factory experiments, neutrino oscillations, neutrinoless double beta decay, grand unified theories, and leptogenesis. 
Moreover, it promotes a systematic understanding of rephasing invariants and is expected to provide a new foundation for the analysis of CP phases. 
The intrinsic simplicity of the Kobayashi--Maskawa parameterization thus serves as a powerful tool for advancing our understanding of CP violation to a deeper level.

\section*{Acknowledgment}

The study is partly supported by the MEXT Leading Initiative for Excellent Young Researchers Grant Number JP2023L0013.


\end{document}